\documentclass[journal]{IEEEtran}

\usepackage{amssymb}
\usepackage{amsfonts}
\usepackage{latexsym}
\usepackage[cmex10]{amsmath}
\usepackage[utf8]{inputenc}
\usepackage{color}
\usepackage{multirow}
\usepackage{hhline}

\ifCLASSINFOpdf
  \usepackage[pdftex]{graphicx} 
\else 
  \usepackage[dvips]{graphicx}
\fi

\newcommand{\erfc}{\mathrm{erfc}\,}
\newcommand{\erf}{\mathrm{erf}\,}
\newcommand{\prt}{p(r,t|r_0)}

\newcommand{\prtzero}{p(r,t \rightarrow 0 |r_0)}

\newcommand{\rrn}{r_r}
\newcommand{\rrs}{r_s}

\newcommand{\betaEqn}{\frac{n \rrs + \pi \rrn}{\pi \rrn^2}}

\newcommand{\fhit}{f_{\text{hit}}(t)}
\newcommand{\Fhit}{F_{\text{hit}}(t)}

\newcommand{\Fhitt}[1]{F_{\text{hit}}(#1)}

\newcommand{\FhitsNwoutT}{F_{\text{hit}}^{r_s,n}}

\newcommand{\FhitsN}{F_{\text{hit}}^{r_s,n}(t)}

\newcommand{\FhitsNt}[1]{F_{\text{hit}}^{r_s,n}(#1)}
\newcommand{\FhitsNtt}[2]{F_{\text{hit}}^{r_s,n}(#1,#2)}

\newcommand{\Nrxint}[2]{N_{\text{rx}}(#1,#2)}
\newcommand{\Ntx}{N_{\text{tx}}}

\usepackage[pdftex,dvipsnames]{xcolor}

\title{Effect of Receptor Density and Size on Signal Reception in Molecular Communication via Diffusion with an Absorbing Receiver}

\author{Ali~Akkaya,
        H. Birkan~Yilmaz,
        Chan-Byoung~Chae,~\IEEEmembership{Senior Member,~IEEE}, and~Tuna~Tugcu,~\IEEEmembership{Member,~IEEE}
\thanks{Manuscript received June, 2014. The work of A. Akkaya and T. Tugcu was supported in part  by State Planning Organization (DPT) of Republic of Turkey under the project TAM with the project number 2007K120610, Bogazici University Research Fund (BAP) under grant number 7436, and by Scientific and Technical Research Council of Turkey (TUBITAK) under Grant number 112E011. The work of H. B. Yilmaz and C.-B. Chae was in part funded by the MSIP (Ministry of Science, ICT \& Future Planning), Korea, under the ``IT Consilience Creative Program" (NIPA-2014-H0201-14-1002) supervised by the NIPA (National IT Industry Promotion Agency) and by the ICT R\&D program of MSIP/IITP.}
\thanks{A. Akkaya and T. Tugcu are with NETLAB, Department of Computer Engineering, Bogazici University, Istanbul, 34342, Turkey (e-mail: ali.akkaya@boun.edu.tr; tugcu@boun.edu.tr)}
\thanks{H. B. Yilmaz and C.-B. Chae are with the School of Integrated Technology,
Yonsei Institute of Convergence Technology, Yonsei University, Korea (e-mail:
birkan.yilmaz@yonsei.ac.kr; cbchae@yonsei.ac.kr)}
}

\begin{document}
\maketitle

\begin{abstract}

The performance of molecular communication is significantly impacted by the reception process of the messenger molecules. The receptors' size and density, however, have yet to be investigated. In this letter, we analyze the effect of receptor density and size on the signal reception of an absorbing receiver with receptors. The results show that, when the total receptor area is the same, better hitting probability is achieved by using a higher number of relatively small receptors. In addition, deploying receptors, which cover a small percentage of the receiver surface, is able to create an effective communication channel that has a detectable signal level.

\end{abstract}


\section{Introduction}

A nanonetwork enables communication between nanomachines; it also bridges nanomachines to higher scale systems \cite{akyildiz2011nanonetworksAN, nakano2012molecularCA}. One viable method for a nano-scale communication approach is found in molecular communication. Such communication, which enables communication between living cells, already exists in nature. It is thus an eligible candidate for inter-nanomachine communication. Molecular Communication via Diffusion (MCvD) is a short-to-medium range molecular communication technique in which the messenger molecules diffuse in the propagation medium to transfer the intended information  \cite{kuran2010energyMF, kim2013novelMT}.

After being released from a transmitter, molecules propagate in their environment by following diffusion dynamics. While most scatter in the environment, some of these molecules, according to their type and the properties of the environment, reach the receiver. In nature, a messenger molecule is received only when it binds to one of the receptors on the surface of the receiver. Then, for most receptor types, the messenger molecules are absorbed by the receiver. Therefore, each molecule contributes to the signal only once due to absorption or other mechanisms. For communication channel design, an important parameter is reception probability. The key factors affecting this parameter are the \emph{size} and the \emph{density} of the receptors. The number of receptors on the cell surface for signaling molecules can vary from $500$ to-for a specific type of messenger molecule-more than $100,000$ per cell [5]. If reception  probability is too low, it may be impossible to establish an efficient communication channel; if it is unnecessarily high, it may denote inefficient use of the resources, specifically in terms of the energy and fragment molecules of the receptor. Reception probability should thus be well analyzed. 

In 1-D and 3-D medium, the hitting rate to a perfectly absorbing spherical receiver is analyzed in \cite{farsad2014channelAN,sirinivas2012molecularCI,yilmaz20143dChannelCF}. In this letter, we formulate the hitting rate of the molecules to the receptors of an absorbing receiver in a 3-D medium and verify our formulation via simulations. We also derive additional formulations to address receiver design issues. First, guidance is provided on the selection of receptor size. We then analyze the total area of the receptors needed to achieve a specific hitting rate. Utilizing these analyses, it is possible to optimize the production costs of the receptors and the receivers.


\section{Molecular Channel Characteristics for Receiver with Receptors}

\subsection{Communication Model}
\begin{figure*}[t]
\centering
 \includegraphics[width=1.0\textwidth,keepaspectratio]{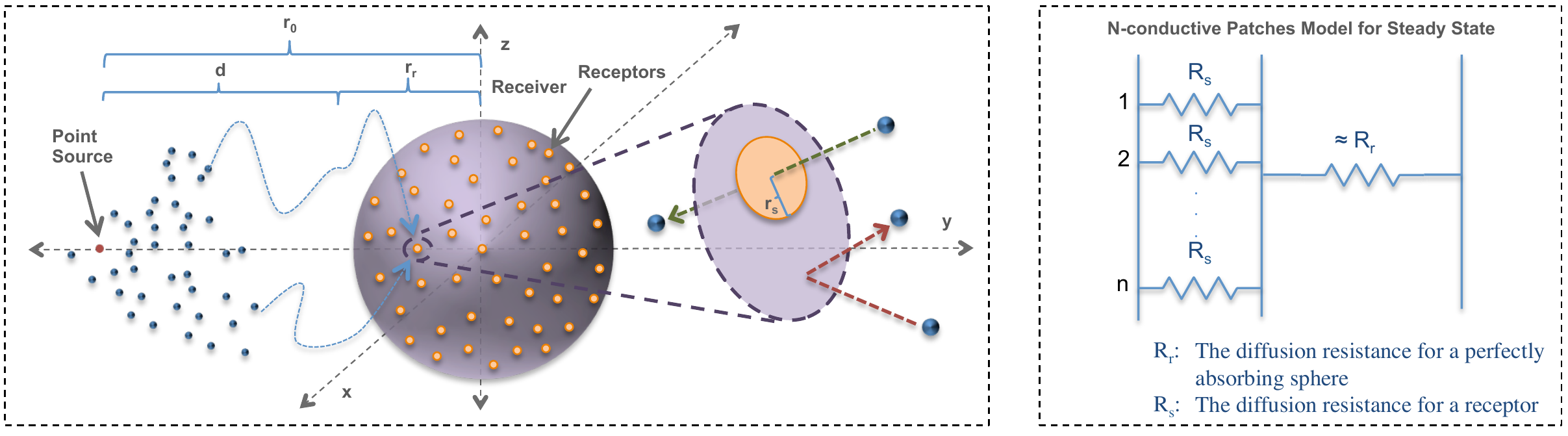}
\caption{Communication model and the equivalent N-conductive patches model.}\label{fig:communication_model}
\end{figure*}


The communication model used in this letter is depicted in Figure  \ref{fig:communication_model}. Messenger molecules are used as the information carriers between a point source and a spherical receiver with absorbing receptors. The point source is located at a distance~$r_0$ from the center of the receiver. The point source and the spherical receiver both reside in a fluid propagation medium. It is assumed that the medium is unconfined, thus extending to infinity in all directions. After the information is modulated onto some physical property of the molecules, the molecules are released to the medium where they diffuse according to Brownian motion and arrive at the receiver. To absorb the molecules, the spherical receiver with radius $r_r$, uses its receptors with radius $r_s$. If a molecule collides with one of the $n$ receptors deployed on the surface of the receiver, it is absorbed by the receiver. If it collides with the surface of the receiver without touching a receptor, it bounces back.

\subsection{Absorption Rate of a Spherical Receiver} \label{spherical_absorber}

The microscopic theory of diffusion roots from the assumption that a substance will move down its concentration gradient. The derivative of the flux with respect to time results in Fick's Second Law in a 3-D environment, given by
\begin{equation}
\label{eq:ficks3}
\frac{\partial \prt}{\partial t} = D \nabla^2 \prt
\end{equation}
\noindent where $\nabla^2$, $\prt$, and $D$ are the Laplacian operator, the molecule distribution function at time $t$ and distance $r$ given the initial distance $r_0$, and the diffusion constant. The value of $D$ depends on the temperature, viscosity of the fluid, and the Stokes' radius of the molecule \cite{tyrrell1984diffusionIL}. 

The fraction of hitting molecules to a spherical receiver located at $(0,0,0)$ can be derived by solving the Fick's diffusion equation with the initial and the boundary conditions obeying the problem and describing the absorbing process following the methodologies in \cite{redner2001guideTF,schulten2000lecturesIT,chaeEL}. 

The initial condition is defined as
\begin{equation}
\prtzero = \frac{1}{4 \pi r_0^2} \delta(r - r_0),
\end{equation}
and the first boundary condition is
\begin{equation}
\label{eq:boundary_inf}
\lim_{r \rightarrow \infty} \prt = 0,
\end{equation}
which reflects the assumption that the distribution of the molecules vanishes at distances far greater than $r_0$. The second boundary condition is
\begin{equation}
\label{eq:boundary_sphere}
D \frac{\partial \prt}{\partial r} = w \,\prt \text{ , for } r = \rrn
\end{equation}
where $\rrn$ and $w$ denote the radius of the receiver and the rate of reaction. Reaction rate with the receiver boundary is controlled by $w$ and $w=0$ means a nonreactive surface while $w$ approaching to infinity corresponds to the boundary in which every collision leads to an absorption. Solving the differential equation for arbitrary $w$ and following the molecule distribution $\prt$, hitting rate of the molecules to the receiver at time $t$ can be obtained as \cite{schulten2000lecturesIT}, 
\begin{multline}
\label{eqn:fhit}
\fhit = \displaystyle\frac{\rrn w}{r_0} \left( \frac{1}{\sqrt{\pi D t}} \exp{\left[ - \frac{(r_0 \!-\! \rrn)^2}{4 D t} \right] } \right. \\  
\left. - \beta \exp{\left[ \beta (r_0 \!-\! \rrn) + \beta^2 D t \right]} \erfc \left[\frac{r_0 \!-\! \rrn}{\sqrt{4Dt}} + \beta \sqrt{Dt} \right]  \right)
\end{multline}
\noindent where $\beta = (w \rrn + D) / (D \rrn)$. Furthermore, by integrating $\fhit$ with respect to time, we can obtain $\Fhit$ for arbitrary $w$, which is the fraction of molecules absorbed by the receiver until time $t$ \cite{schulten2000lecturesIT}:
\begin{multline}
\label{eqn:Fhit}
\Fhit = \displaystyle\frac{\rrn \beta - 1}{r_0 \beta} \left( 1 + \erf\left[ \frac{\rrn - r_0}{\sqrt{4 D t}} \right] \right. \\
\left. - \exp{\left[ (r_0 - \rrn) \beta  + D t \beta^2 \right]} \erfc\left[\frac{r_0 - \rrn + 2 D t \beta}{\sqrt{4Dt}}\right]  \right).
\end{multline}

\subsection{Absorption Rate of a Spherical Receiver with Absorbing Receptors}

In nature, a molecule is received by a receiver only when it binds to one of the receptors on the surface. To abstract this phenomenon, we model the receptors as circular areas over the receiver surface. A diffusing molecule is absorbed by the receiver only when it collides with a receptor. The other parts of the receiver surface are not capable of absorbing molecules. To model such a receiver, we need to derive the special case of (\ref{eqn:Fhit}), where $w$ depends on the number of receptors $n$, and the radius of receptors, $\rrs$.

We start by investigating the boundary condition for (\ref{eqn:Fhit}) as ${t\rightarrow \infty}$ , which gives us the fraction of received molecules for arbitrary $w$ and steady state
\begin{align}
\begin{split}
\label{eqn:FhitInfinity}
\lim_{t \rightarrow \infty} \Fhit =  \displaystyle\frac{\rrn \beta - 1}{r_0 \beta}.
\end{split}
\end{align}

We can also formulate the fraction of molecules absorbed when $t\rightarrow \infty$ using an analogy with the electricity domain where $n$ conductive patches are located on an insulating sphere, assuming $r_s \ll r_r$ \cite{berg1993randomWI}. The insulating sphere is analogous to the receiver and the receptors that bind with the molecules are analogous to the patches through which the current flows. For this scenario, the diffusion current $I$, which corresponds to the current in the electricity domain, is given by
\begin{equation}
\label{eqn:DiffusionCurrent}
I  = C / R
\end{equation}
\noindent where $C$ is the concentration difference and $R$ is the diffusion resistance. The diffusion resistance for a sphere with absorbing receptors $R$ can be written as

\begin{equation}
\label{eqn:DiffusionRessistance}
R = R_r \left(1 + \frac{\pi  r_r }{ n r_s}\right)
\end{equation}
\noindent where $R_r$ is the diffusion resistance of a perfectly absorbing sphere \cite{berg1993randomWI}. This equation shows that the diffusion resistance of a receiver with receptors is larger than that of a perfectly absorbing sphere by a factor of 
$1 + (\pi r_r) / (n r_s$). Using (\ref{eqn:DiffusionCurrent}) and (\ref{eqn:DiffusionRessistance}), for the steady state, we can write 
\begin{equation}
\label{eqn:DiffusionCurrentRatio}
\frac{I}{I_r} = \frac{1}{ 1 + \frac{\pi r_r }{ n r_s}}
\end{equation}
\noindent where $I$ and $I_r$ are the diffusion current for a sphere with absorbing receptors and a perfectly absorbing sphere~\cite{berg1993randomWI}. The fraction of molecules absorbed for an absorbing sphere at ${t \rightarrow \infty}$ is $\rrn/r_0$, hence we can write the fraction of molecules for a sphere with absorbing receptors as
%
\begin{equation}
\label{eqn:FhitsnInfinity}
\lim_{t \rightarrow \infty} \FhitsNt{t} = \frac{\rrn}{r_0}\frac{1}{ 1 + \frac{\pi r_r }{ n r_s}} = \frac{\rrn}{r_0}\frac{\rrs n}{\rrs n + \pi \rrn}
\end{equation}

At the boundary condition where $t\rightarrow \infty$, (\ref{eqn:FhitInfinity}) and (\ref{eqn:FhitsnInfinity}) will be equal. Using this equality, we can write $w$ and $\beta$ as,

\begin{figure}[t]
    \centering
    \includegraphics[width=1.0\columnwidth,keepaspectratio]{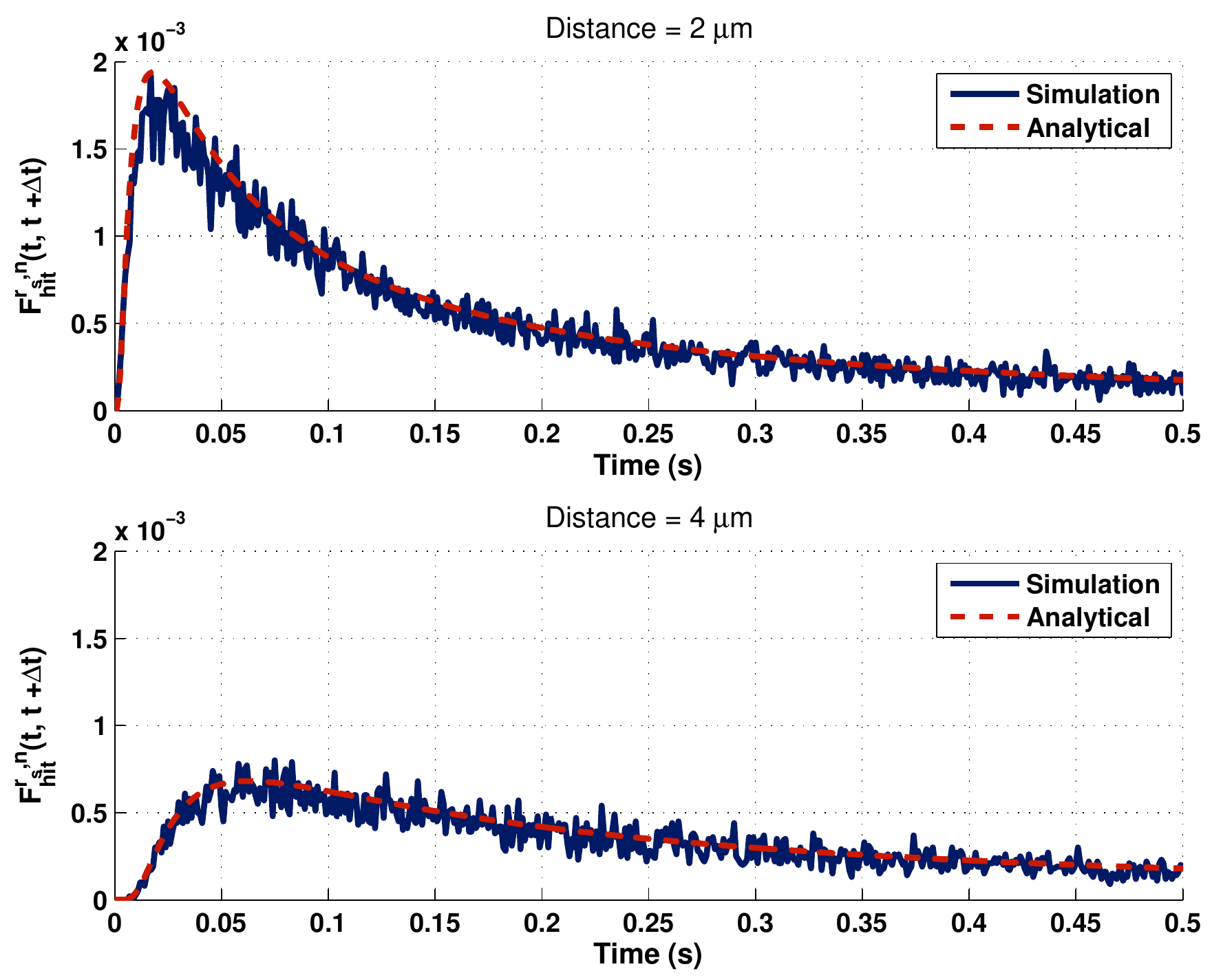}
    \caption{Analytical and simulation results for the fraction of molecules absorbed by the receiver versus time for different $d$ values ($r_r=10 \mu m$, $r_s=0.04 \mu m$, $n=1250$, $D=79.4 \mu m^2/s$, $\Delta t=0.001\,$s, the number of simulations $=100000$). }
     \label{fig:TimeVsPhit}
\end{figure}

\begin{equation}
\label{eqn:w}
w = \frac{n \rrs D}{\pi \rrn^2},~~\beta = \betaEqn
\end{equation}
%
%
Using (\ref{eqn:w}) and (\ref{eqn:Fhit}), we can derive the formula for a receiver with absorbing receptors as, 
\begin{multline}
\label{eqn:Fhitsn}
\FhitsN = \displaystyle\frac{\rrn}{r_0}\frac{\rrs n}{\rrs n + \pi \rrn}  \left( 1 + \erf\left[ \frac{\rrn - r_0}{\sqrt{4 D t}} \right] \right. \\
\left. - \exp{\left[ (r_0 - \rrn) \left(\betaEqn\right)  + D t \left(\betaEqn\right)^2 \right]} \right. \\
\left. \times\erfc\left[\frac{r_0 - \rrn + 2 D t (\betaEqn)}{\sqrt{4Dt}}\right]  \right).
\end{multline}

\noindent Using \eqref{eqn:Fhitsn}, we define the fraction of molecules received between $t_1$ and $t_2$ as
\begin{align}
\label{eqn:Fhitsnt}
\FhitsNtt{t_1}{t_2} = \FhitsNt{t_2}-\FhitsNt{t_1}.
\end{align}

As shown in Figure \ref{fig:TimeVsPhit}, the simulation results for two different distances are inline with the analytical formulation given in (\ref{eqn:Fhitsnt}). As expected, for shorter distances, the amplitude of the signal is higher and the signal peak is observed earlier.

\subsection{The Number of Received Molecules}
The expected number of molecules hitting the receiver in the interval $[t_1, t_2]$ for a given number of receptors can be evaluated by 
\begin{equation}
\label{eqn:NrxInterval}
\mathbb{E}[\Nrxint{t_1}{t_2}]  = \Ntx \,\,\FhitsNtt{t_1}{t_2},
\end{equation}
\noindent where $\Ntx$ denotes the number of emitted molecules at $t=0$. The signal at a desired resolution, $\Delta t$, can be easily obtained by plotting the expected number of received molecules.

\begin{figure}[t]
    \centering
    \includegraphics[width=1.0\columnwidth,keepaspectratio]{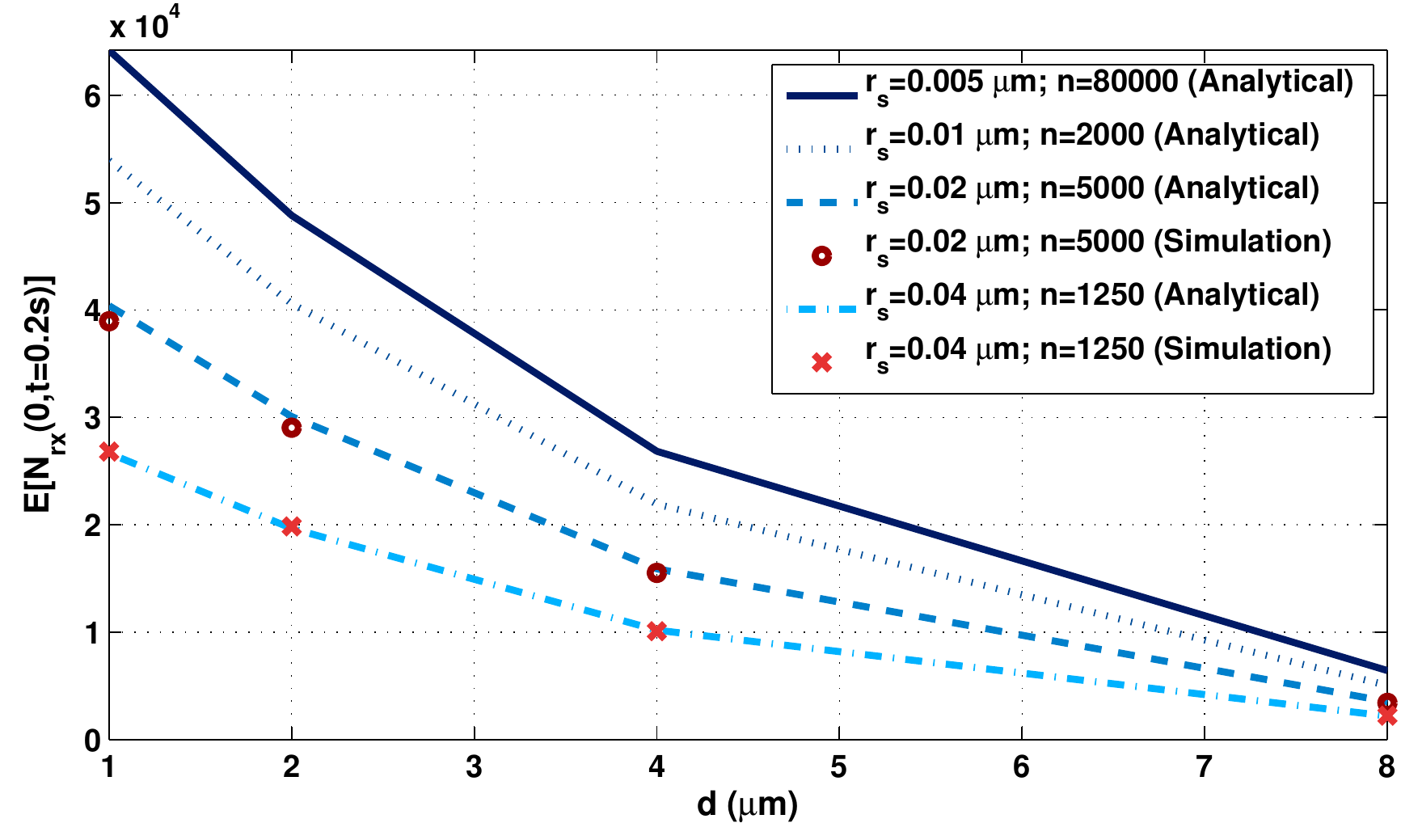}
    \caption{The number of received molecules till $t=0.2s$ versus the distance for the same total receptor area ($r_r=10\, \mu m$, total receptor area = 2$\pi$, ${\Ntx=100000}$ molecules, $D=79.4\, \mu m^2/s$).}
     \label{fig:DistanceVsPhitPulsePeak}
\end{figure}

Figure \ref{fig:DistanceVsPhitPulsePeak} shows the number of received molecules until $t=0.2 s$ versus distance $d$ where the total area that is covered by the receptors is kept constant for different $r_s$ and $n$ values. The $r_s=\{0.04, \, 0.02\} \,\mu m$ cases are simulated and the simulation results are coherent with the analytical results. The results indicate guidance for an important receptor design criteria. For any fixed distance, given the total area that will be covered by the receptors, to achieve a better hitting probability, one should use a higher number of relatively small receptors. The achieved nominal gain gets smaller as the distance increases.

\subsection{Receptor Area Analysis}

To create an efficient communication channel, once the appropriate receptor type has been selected, it is important to find the minimum sufficient ratio of the total surface area that should be covered with receptors. This decision has a direct effect on receiver production costs. 

\begin{figure}[t]
    \centering
    \includegraphics[width=1.0\columnwidth,keepaspectratio]{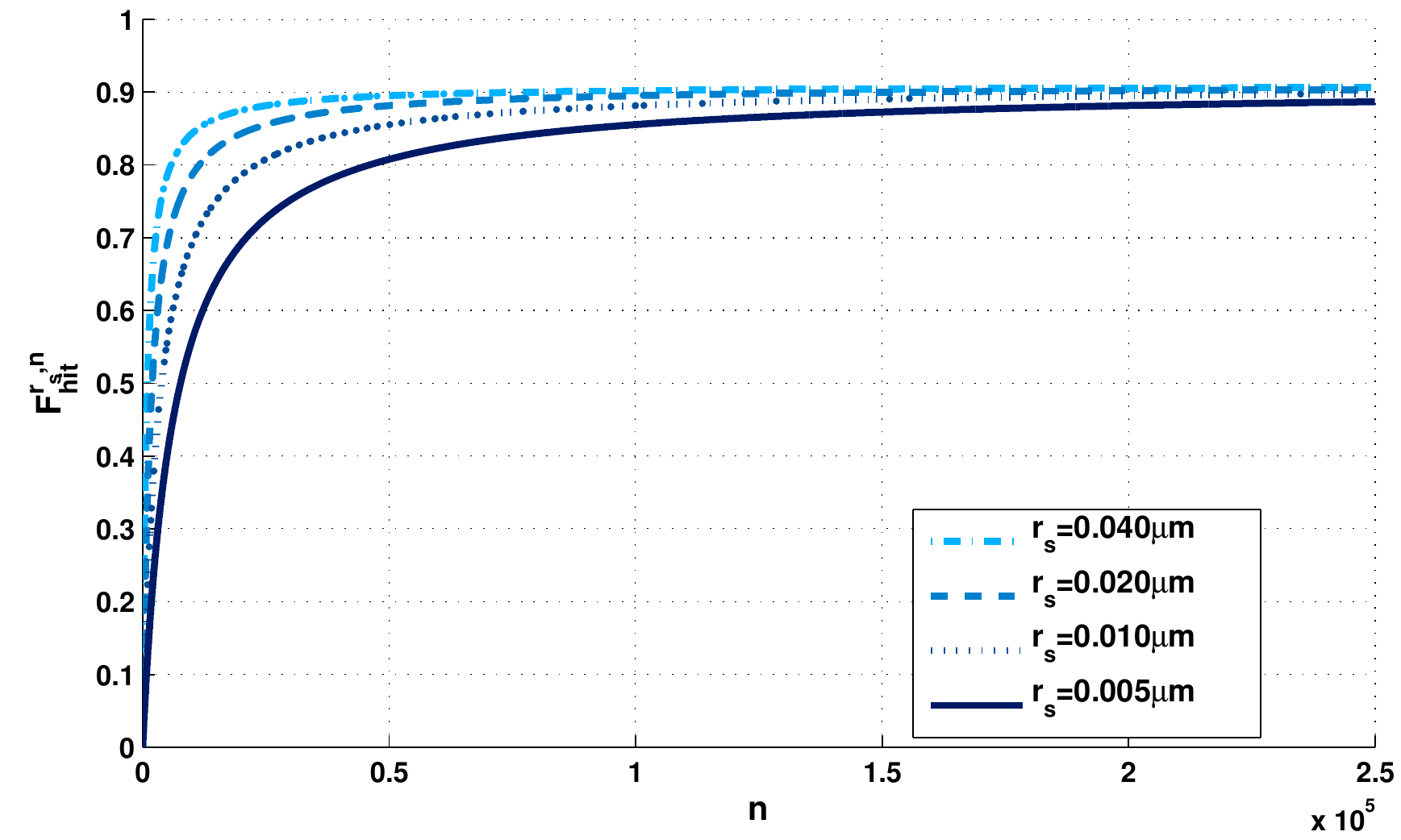}
    \caption{Fraction of molecules absorbed by the receiver versus the number of receptors for different $r_s$ values ($r_r=10 \mu m$, $r_0=11 \mu m$, $t \rightarrow \infty$, $D=79.4 \mu m^2/s$). }
     \label{fig:NVsPhit}
\end{figure}

To analyze this, we should have a formula or a method to find the minimum number of receptors needed to achieve a specific $\FhitsN$ value, $\alpha$, for the given parameters, which is denoted as $n_{\alpha}$. Note that, when $r_0$, $r_r$, $r_s$, $D$, $t$ are fixed, $\FhitsNwoutT$ strictly increases as the number of receptors increase as shown in Figure \ref{fig:NVsPhit}. Therefore, for a given $\alpha$, the set $\{n | \FhitsNwoutT=\alpha\}$ has a single element, which is $n_{\alpha}$. Since $\FhitsNwoutT$ strictly increases with respect to $n$, it enables us to perform a numerical search using (\ref{eqn:Fhitsn}) to find $n_{\alpha}$.

The ratio of the total area of the receptors to the total area of the perfectly absorbing receiver can be written as
\begin{equation}
\label{eqn:receptorAreaRatio}
\frac{A^{r_s,n}}{A} = n \left(\frac{ r_s  }{ 2 r_r}\right)^2
\end{equation}
where $A^{r_s,n}$ is the surface area covered by the receptors and $A$ is the total surface area of the receiver. Finally, using $n_{\alpha}$ and (\ref{eqn:receptorAreaRatio}), we can calculate the ratio of the surface area to be covered by receptors to achieve $\FhitsNwoutT = \alpha$ using

\begin{equation}
\label{eqn:Aat}
A_{\alpha} = n_{\alpha} \left(\frac{ r_s  }{ 2 r_r}\right)^2.
\end{equation}

\begin{figure}[t]
    \centering
    \includegraphics[width=1.0\columnwidth,keepaspectratio]{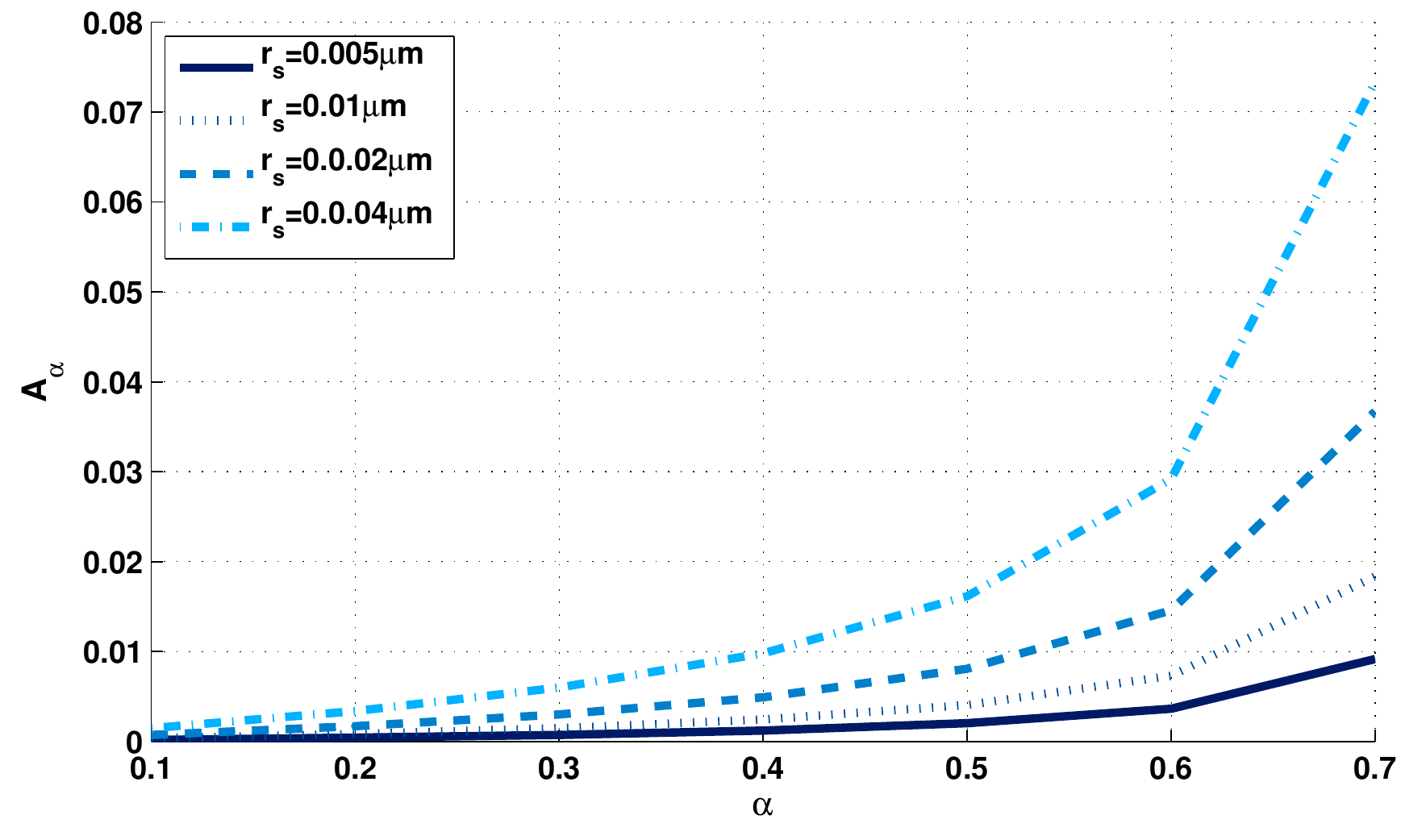}
    \caption{Ratio of the surface area to be covered by receptors to achieve $\FhitsNt{t} = \alpha$ ($r_r=10 \mu m$, $r_0=11 \mu m$, $D=79.4 \mu m^2/s$, $t=0.2 \,s$).}
     \label{fig:AreaVsAlpha}
\end{figure}

Figure \ref{fig:AreaVsAlpha} illustrates $A_\alpha$ versus $\alpha$ for $t=0.2s$. The results show that, to achieve significant $\FhitsNt{t}$ values, it is sufficient to deploy receptors to cover only a miniscule ratio of the total surface area. For instance, for $r_s=0.005\,\mu m$, the ratio of the receptor area to the full surface area so as to achieve $\alpha=0.7$ is 0.0092. Hence, it is possible to achieve ${\FhitsNt{t}=0.7}$ by covering less than 1\% of the total surface area of the receiver, where $\Fhitt{t}=0.7811$ for perfectly absorbing sphere. This shows that it is possible, on a practical level, to deploy receptors for several different molecule types and achieve considerable signal energy for each communication channel that uses different molecule types.

\section{Conclusion}

In this letter, we considered the imperfect reception process in nature to build a more realistic model and derive an analytical formulation of the absorption rate of a spherical receiver with absorbing receptors in a 3-D medium. We also addressed  receiver design issues, specifically, the optimization of the size and the total area of the receptors. This has a direct effect on the production costs of the receptors and receivers. We used the formulations to conclude that it is possible, at a practical level, to have a comparable signal energy to a perfectly absorbing sphere while covering as little as 1\% of the surface area of the receiver. We will consider multiple pairs of receptor-messenger molecules and conduct energy optimization in our future work.

\bibliographystyle{IEEEtran}
\bibliography{IEEEabrv,nanoHuge}

\end{document}